\documentclass[reprint,aps,pra,twocolumn,longbibliography]{revtex4-1}

\usepackage{amsmath, amsthm, amssymb,braket,nicefrac}
\usepackage[usenames,dvipsnames]{xcolor}
\usepackage{lmodern,microtype,ragged2e, comment}
\usepackage{appendix,hyphenat,mathtools,dsfont,bm,bbm}
\usepackage[makeroom]{cancel}
\usepackage{graphicx}
\usepackage{soul}

\definecolor{mygrey}{gray}{0.35}
\definecolor{myblue}{rgb}{0.2,0.2,0.8}
\definecolor{myzard}{cmyk}{0,0,0.05,0}
\definecolor{mywhite}{rgb}{1,1,1}
\definecolor{myred}{rgb}{0.9,0.1,0.}
\definecolor{dgreen}{rgb}{0.0, 0.5, 0.0}
\usepackage[colorlinks=true,citecolor=myblue,linkcolor=myblue,urlcolor=myblue]{hyperref}
\usepackage{cleveref}

\DeclareGraphicsExtensions{.png,.pdf,.eps,.jpg}
\graphicspath{ {./figs/} }
\newcommand{\proj}[1]{\ket{#1}\!\bra{#1}}
\newcommand{\ketbra}[2]{\left|#1\rangle\langle#2\right|}
\newcommand{\inner}[2]{\left\langle #1\lvert#2\right\rangle}
\newcommand{\abs}[1]{\left\lvert #1\right\rvert}
\newcommand{\be}{\begin{equation}} 							
\newcommand{\ee}{\end{equation}}
\newcommand{\bematrix}{\left(\begin{matrix}}
\newcommand{\ematrix}{\end{matrix}\right)}
\def\one{{\mbox{$1 \hspace{-1.0mm}  {\bf l}$}}}						

\def\R{{\ensuremath{\mathbbm R}}}

\def\ii{\mathrm{i}}
\def\tr{\mathrm{tr}}
\def\d{\mathrm{d}}

\def\paulivec{\bm \sigma}

\def\sx{\sigma_x}
\def\sy{\sigma_y}
\def\sz{\sigma_z}

\def\cB{\mathcal B}

\def\cD{\mathcal D}

\def\cF{\mathcal F}
\def\cG{\mathcal G}
\def\cH{\mathcal H}											
\def\cI{\mathcal I}
\def\cJ{\mathcal J}

\def\cO{\mathcal O}
\def\cP{\mathcal P}

\def\cR{\mathcal R}

\def\cU{\mathcal U}

\newcommand\defn[1]{\textsl{#1}}

\newcommand{\eqnref}[1]{Eq.~(\ref{#1})}
\newcommand{\eqnsref}[2]{Eqs.~(\ref{#1}) and (\ref{#2})}
\newcommand{\figref}[1]{Fig.~\ref{#1}}

\newcommand{\secref}[1]{Sec.~\ref{#1}}
\newcommand{\appref}[1]{App.~\ref{#1}}

\newcommand{\citeref}[1]{Ref.~\cite{#1}}

\begin{document}

\title{Super-resolving collective quantum measurements}
\author{J.O. de Almeida$^{1,2}$}
\email{jessica.almeida@itp3.uni-stuttgart.de}
\author{M. Lewenstein$^{2,3}$}
\author{M. Skotiniotis$^{4, 5, 6}$}%
\affiliation{$^{1}$Institute for Theoretical Physics III, University of Stuttgart, 70550 Stuttgart, Germany\\
$^{2}$ICFO - Institut de Ciencies Fotoniques, The Barcelona Institute of Science and Technology, Av. Carl Friedrich Gauss 3, 08860 Castelldefels (Barcelona), Spain\\
$^{3}$ICREA, Pg. Llu\'is Companys 23, 08010 Barcelona, Spain\\
$^{4}$F\'isica Te\'orica: Informaci\'o i Fen\`omens Qu\'antics, Departament de Fisica, Universitat Aut\`onoma de Barcelona, E-08193, Bellaterra
(Barcelona), Spain\\
$^{5}$Quantum Thermodynamics and Computation Group, Department of Electromagnetism and Condensed Matter, Universidad de Granada, 18010 Granada, Spain\\
$^{6}$Institute Carlos I for Theoretical and Computational Physics, Universidad de Granada, 18010 Granada, Spain}

\begin{abstract}

We demonstrate a method for super-resolving signals encoded as finite mixtures of bosonic modes using collective measurements that exploit permutation symmetry. 
Specifically, we use multiple copies of the state $\rho$ of the finite mixture to extract an estimate for the purity of $\rho$ via a spectrum 
measurement---the weak Schur-sampling measurement.  Depending on the outcome we then further fine-grain the measurement to optimally extract
an estimate of the relative intensity between the two incoherent mixtures. Our protocol furnishes simultaneous estimates for both the relative intensity and the 
separation of incoherent signals saturating the multi-parameter Cram\'{e}r-Rao bound, and is robust against misalignment errors.  We also provide viable experimental 
avenues for implementing such collective measurements in different set-ups.
\end{abstract}

\date{\today}

\maketitle

\section{\label{sec:Introduction} Introduction}
    	
	The ability to resolve two or more signals is of fundamental importance in a variety of tasks, particularly in communication~\cite{Gisin07} and 
	imaging~\cite{Moreau19}.  For instance, the capacity of a communication channel depends on how well one can resolve the various signals sent through the 
	channel~\cite{Hausladen96}, while the resolving power of telescopes~\cite{Labeyrie77} and microscopes~\cite{Ram06,Cremer13} is defined in terms of 
	their ability to distinguish between two or more objects~\cite{Rayleigh80}.  Regardless of the signal parameters to be resolved are (position, momentum, 
	frequency, or time), resolution limits can be imposed on any measurement seeking to gain information about these parameters through statistical inference 
	methods.  In optical imaging, for example, the resolution of intensity measurements is dictated by Rayleigh's criterion~\cite{Rayleigh80}, and whilst sub-Rayleigh 
	resolution is possible, it comes at the cost of a decreasing signal-to-noise ratio~\cite{Shahram04}.

    	In recent years techniques from quantum statistical inference have shown that such heuristic limits can be overcome. For the case of imaging light from two equally 
    	bright, incoherent point sources~\citeref{Tsang16} derived a Spatial Mode Demultiplexing (SPADE) measurement capable of resolving the two incoherent sources 
	regardless of their separation. Such super-resolving measurements have been the subject of increasing interest both 
	theoretically~\cite{Nair16,Lupo16,Ang17,Rehacek17,Rehacek18,Lu18,Napoli19,Prasad19,Tsang19,Lupo20,Datta20,Hradil21,Liang21,Karuseichyk22} 
    	and experimentally~\cite{Yang16,Tham17,Yang17,Donohue18,Zhou19,Boucher20,Wadood21}.  However, super-resolving measurements processing the incoming light 
    	photon-by-photon rely on a priori knowledge of the position of the intensity centroid of the imaging systems Point Spread Function (PSF). Whilst such information 
	can be estimated by sacrificing part of the signal~\cite{Grace20}, it is known that any misalignment with respect to the intensity centroid severely affects the 
	measurement's super-resolving power~\cite{Chrostowski2017, Gessner20, deAlmeida21,Stan2023}. Additionally, super-resolving measurements are extremely 
	sensitive to the presence of noise~\cite{Bonsma-Fisher19,Gessner20,Len20,Santamaria24} and are incompatible in the multi-parameter case where one wishes to 
	estimate both the separation as well as the relative intensity of the two sources~\cite{Rehacek17,Rehacek18,Shao22,Santamaria24}. This poses a significant 
	challenge in scenarios where the signal to be imaged is limited---such as in exoplanet detection~\cite{Huang21,Huang23}---preventing the estimation of each 
	parameter separately. Such strategies are also sub-optimal as they do not provide any information about the correlations between the various parameters.  
	
	Here, we address the problem of estimating both the relative intensity $q$ and the separation, $\epsilon$, of two incoherent sources when their separation falls 
	below the Rayleigh limit~\cite{Rayleigh80}. Our protocol exploits the inherent permutation symmetry of multiple copies of the mixed state $\rho$, to provide an 
	estimate of the laters purity.  This is achieved by the weak Schur-sampling measurement~\cite{Childs07} (henceforth referred t as the \defn{spectrum} 
	measurement), which is robust against misalignment.  As the purity is related to both the relative intensity and separation
	of the signals, its estimation enables inference of the separation provided we have an estimate of the relative intensity. The latter can be obtained by linear 
	post-processing of the spectrum measurement via a unitary operation (which depends on the value of the purity) followed by a projective measurement.
	We demonstrate that our protocol remains valid both when the intensity centroid is known as well as under arbitrary and unknown misalignment provided the 
	latter lies within the span of the two signals. This collective measurement strategy achieves the smallest possible covariance matrix for the simultaneous estimation 
	of $\epsilon$ and $q$, in the limit of large sample sizes.

	Our measurement strategy generalizes the Hong-Ou-Mandel effect\cite{HOM}, which has been shown to achieve sub-Rayleigh resolution in 
	interferometry~\cite{Lyons18, Yepiz-Graciano20, Aguilar20} as well as in spatial~\cite{Parniak18, Ndagano22, Gao22}, temporal~\cite{Ansari21}, and 
	spectral~\cite{Prakash2021,Mazelanik22} imaging.  As finite mixtures of highly overlapping signals frequently arise in signal 
	processing~\cite{Hyvarinen2009,Stogbauer2004} and spectroscopy~\cite{Monakhova2010}, our results demonstrate that both the separation and 
	relative intensity of such signals can be optimally estimated at the fundamental limit using a practical collective measurement strategy.
	    
    	The outline of the paper is as follows. In~\secref{sec:intro} we formulate the quantum analog of two incoherent bosonic sources producing a two-component 
	mixture of highly overlapping signals (\secref{sec:model}) and introduce a mapping of the problem into the simplest two-dimensional quantum system---the 
	qubit (\secref{sec:qubit_model}).  We also review the necessary statistical inference framework in \secref{sec:CramerRao}. In \secref{sec:QFI}, we derive 
	analytical expressions for the Quantum Fisher Information Matrix (QFIM) and Symmetric Logarithmic Derivatives (SLDs) to estimate the purity, relative intensity, 
	and orientation of the bosonic quantum state, both in the presence and absence of misalignment. We further show how these quantities lead to optimal 
	estimates for the separation and relative intensity of the two-component signal.  In \secref{sec:collective} we exploit permutation symmetry to construct a 
	collective measurement strategy for optimally estimating the purity and relative intensity of the signals given $N$, identically prepared bosons. In 
	\secref{sec:Implementation} we discuss possible implementations of such collective measurements both in photonic as well as atomic setups.  
	We summarize our results in \secref{sec:Conclusions}.

\section{\label{sec:intro}Preliminaries}

    	In this section we give a formal quantum mechanical description of the problem of resolving a two-component mixture of highly overlapping signals encoded in a 
	single-mode bosonic quantum state (\secref{sec:model}).  We show how the problem can be mapped to a two-dimensional subspace, $\cH_2$, where the 
	pertinent information regarding the state of the system are encoded in the Bloch vector of a positive semi-definite operator of unit trace, 
	$\rho\in\cB(\cH_2)$~\cite{Chrostowski2017} (\secref{sec:qubit_model}), and frame the problem of determining the separation and relative weights of 
	the two signals in terms of quantum multi-parameter estimation theory, which we review in \secref{sec:CramerRao}.

	\subsection{\label{sec:model} Two-component mixtures carried by a single bosonic mode}
	
	Consider a signal encoded in the degrees of freedom of a single bosonic system whose state space is $\cH$.  The latter can be finite or infinite dimensional 
	depending on the degrees of freedom used to carry the signal.  For instance if the signal is encoded in the boson's polarization then $\cH\cong\mathbb{C}^2$, 
	whereas if the signal is encoded in the bosons position then $\cH\cong L^2(\R)$.  Now let $\ket{\psi},\ket{\phi}\in\cH$ denote the quantum states encoding two 
	signals with $c=\abs{\inner{\psi}{\phi}}\in(0,1)$ denoting their overlap.  A two-component finite mixture of the signals is represented by the density matrix, 
	$\rho \in\cB(\cH)$,
		\be
			\rho = q \ketbra{\psi}{\psi} + (1-q) \ketbra{\phi}{\phi}\, ,
		\label{eq:finite_mixture}
		\ee   
	a positive linear operator, acting on the state space $\cH$, with unit trace.  The parameter $q\in(0,1)$ represents the relative intensity of the signals.  
	Given the state in~\eqnref{eq:finite_mixture}, the goal is to estimate the parameters, $c$ and $q$.    
	
	A particular realization of \eqnref{eq:finite_mixture}, with applications in communication and imaging, is when the signals are encoded using Gaussian states of light.
	For ease of exposition, let $x\in \mathbb{R}$ denote the position degree of freedom of a single photon, although we stress that what follows holds equally well for 
	other continuous degrees of freedom such as momentum, frequency or time.  Moreover, position is the relevant degree of freedom for quantum imaging where the 
	phenomenon of super-resolving quantum measurements first appeared, and we adopt the standard approach for modelling the single-photon state of 
	two incoherent point sources in a diffraction-limited imaging system as an incoherent mixture~\cite{Tsang16}. Denoting by $\{\ket{x}:=a^\dagger(x)\ket{0}\}$, 
	where $\ket{0}$ is the vacuum and $a^\dagger(x),\, a(x)$ are the bosonic creation and annihilation operators satisfying $[a(x),a^\dagger(x')]=\delta(x-x')$, 
	the position eigenbasis any state $\ket{\psi}\in L^2(\mathbb{R})$ can be expressed as 
		\be
			\ket{\psi}=\int_{-\infty}^\infty \d x\, \psi(x)\ket{x}\, .
		\label{eq:singals}
		\ee
	The probability of observing a boson at position $x\in \mathbb{R}$ is given by the square of the absolute value of its wavefunction, i.e., $p(x)=\abs{\psi(x)}^2$.
	The overlap between the two signals is given by 
		\be
			c = \abs{\int_{-\infty}^\infty \d x\, \psi^*(x)\phi(x)}\, .
		\label{eq:overlap}
		\ee 
	
\begin{figure}[ht!]  
 \centering
   \includegraphics[width=0.48\textwidth]{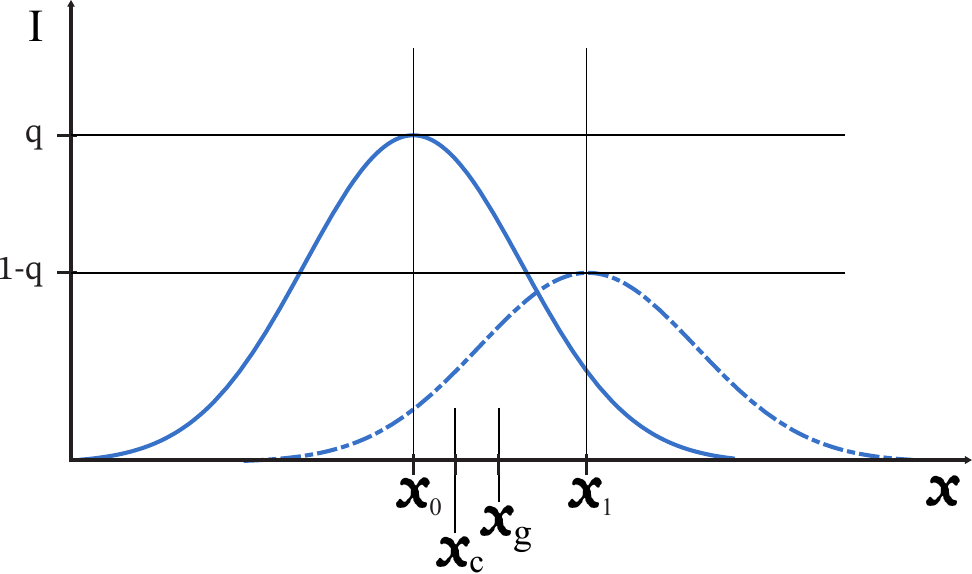}
    \caption{The distribution of intensity as a function of the one-dimensional spatial coordinate for bosons described by the state $\rho$ of \eqnref{eq:finite_mixture} with $\ket{\psi},\,\ket{\phi}\in L^2(\mathbb{R})$ given by \eqnref{eq:Gaussian}.  The separation between the two signals is $\epsilon=x_1-x_0$. The geometric centre of the two sources is $x_g=x_0+\frac{\epsilon}{2}$ whilst the centre of intensity is $x_c=x_0+\epsilon(1-q)$.  For $\frac{\epsilon}{\sigma}\geq 1$ the sources can be resolved by measuring the intensity along the $x$-axis (direct imaging). Sub-Rayleigh and super-resolving measurements deal with the regime $\sigma\gg\epsilon$, i.e., for distributions with very high overlap.}  
   \label{fig:2sources}
\end{figure}

	For the case of imaging the states  $\ket{\psi},\,\ket{\phi}\in L^2(\mathbb{R})$ correspond to the quantum mechanical wavefunctions of a photon emitted by a 
	pair of faint, incoherent, quasi-monochromatic, point sources and transmitted through a one-dimensional, translationally-invariant optical imaging system of unit 
	magnification.  Specifically, $\abs{\psi(x)}^2,\,\abs{\phi(x)}^2$ correspond to the PSFs of the imaging system for the two point sources~\cite{Kurdzialek2022}.  
	The exact form of these PSFs depends crucially on the size and shape of the aperture of the imaging system.  For a suitably masked circular aperture the PSF of an 
	incoherent point source can be modified to a Gaussian
		\be
			\abs{\psi(x)}^2=\frac{1}{\sqrt{2\pi\sigma}}e^{-\frac{(x_0-x)^2}{2\sigma^2}}\, ,
		\label{eq:Gaussian}
		\ee
	whose mean intensity is at $x_0$ and whose spread is $\sigma^2$.  This PSF describes the distribution of intensity at the image plane of the device throughout 
	the entire observation time.  Assuming the sources are faint, so that at most a single photon is transmitted through the image system at any given time, 
	\eqnref{eq:Gaussian}
	also represents the probability of finding a single photon at position $x\in\mathbb{R}$  on the screen of the imaging system.  Hence, the state 
	$\ket{\psi}\in L^2(\mathbb{R})$ is given by  
	
		\be
			\ket{\psi}\equiv\ket{\psi_0(x_0)}=\int_{-\infty}^{\infty}\, \frac{1}{\sqrt[4]{2\pi}}e^{-\frac{(x_0-x)^2}{4}}\, \ket{x}\d x\, ,
		\label{eq:source_left}
		\ee  
	where we have assumed, without loss of generality, that $\sigma^2=1$ and we have taken the real and positive square root of \eqnref{eq:Gaussian}.  Using similar 
	reasoning the state $\ket{\phi}\in L^2(\mathbb{R})$ is given by $\ket{\psi_0(x_0+\epsilon)}$ (see~\figref{fig:2sources}).
	
	Hence, when imaging two faint, quasi-monochromatic, incoherent point sources the state of a single photon is defined by the density operator of  
	\eqnref{eq:finite_mixture} with $\ket{\psi}, \ket{\phi}\in L^2(\mathbb{R})$ given as in \eqnref{eq:Gaussian}.  
\subsection{\label{sec:qubit_model} The qubit model}

	The description above can be significantly simplified to one where the relevant state space associated with their quantum mechanical state of 
	\eqnref{eq:finite_mixture} is that of a finite two-dimensional quantum system.  Specifically,  consider the two-dimensional subspace spanned by 
	$\{\ket{\psi_0(x_0)},\, \ket{\psi_0^\perp(x_0)}\}$, where $\ket{\psi_0^\perp(x_0)}$ is a state in the subspace orthogonal to the fundamental Hermite-Gauss 
	mode centred at $x_0$.  
	
	In the qubit model the distance between the two sources $\epsilon$ is much smaller that the sources width $\sigma$, in this regime a translation in position corresponds to a unitary operation in the Hermite-Gauss basis~\cite{Santos25} and the two sources state of \eqnref{eq:finite_mixture} can be written in the Bloch form.
	
	The state the state $\ket{\psi_0(x_0+\epsilon)}\in L^2(\R)$ in the Hermite-Gaussian basis is
	%With respect to this basis the state $\ket{\psi_0(x_0+\epsilon)}\in L^2(\R)$ can be written as 
		\be
			\ket{\psi_0(x_0+\epsilon)}=c\, \ket{\psi_0(x_0)}+\sqrt{1-c^2} \ket{\psi_0^\perp(x_0)}\, .
			\label{eq:qubit_expand}
		\ee	
	The state $\rho$ in \eqnref{eq:finite_mixture} has support over a two-dimensional subspace spanned by $\{\ket{\psi_0(x_0)},\, \ket{\psi_0^\perp(x_0)}\}$ 
	and can be written in Bloch form as 		
		\be
			\rho=\frac{\one + {\bf r}\cdot\paulivec}{2},
			\label{eq:Bloch_form}
		\ee
	where $\paulivec=(\sx,\sy,\sz)^T$ is the vector of Pauli matrices defined with respect to the orthonormal basis 
	$\{\ket{\psi_0(x_0)},\, \ket{\psi_0^\perp(x_0)}\}$.   The Bloch vector ${\bf r}\in \R_3$ is given by
		\be
			{\bf r}=\bematrix
						2c\sqrt{1-c^2}(1-q)\\
						0 \\
						2q-1 +2c^2(1-q)
					\ematrix = r \,\hat{\bf r}\, ,
		\label{eq:Blochvector}
		\ee
	with
		\be
			r=\sqrt{1-4q(1-q)(1-c^2)}\, ,
		\label{eq:purity}
		\ee
	the purity of $\rho$, whereas the unit vector $\hat{\bf r}=(\sin\theta,\,0,\,\cos\theta)^T$ defines its orientation. The angle 
		\be
			\theta = \cos^{-1}\left(\frac{2q-1 +2c^2(1-q)}{r}\right)
		\label{eq:angle}
		\ee
	specifies the direction of $\rho$ relative to the orthonormal basis $\{\ket{\psi_0(x_0)},\, \ket{\psi_0^\perp(x_0)}\}$, aligned with the $\hat{\bf z}$-axis 
	without loss of generality. The eigenvalues of $\rho$ are  
		\be
			\lambda_{\pm}=\frac{1\pm r}{2}\, ,
		\label{eq:eigenvalues}
		\ee
	and the corresponding eigenvectors are given by    
		\be
			\ket{\lambda_{\pm}}=\sqrt{\frac{1\pm\cos\theta}{2}} \ket{\psi_0(x_0)} \pm \sqrt{\frac{1\mp\cos\theta}{2}}  \ket{\psi_0^\perp(x_0)}\,.
		\label{eq:eigenstates}
		\ee
	In terms of the non-orthonormal basis $\{\ket{\psi_0(x_0)},\, \ket{\psi_0(x_0+\epsilon)}\}$ the eigenvectors of $\rho$ are given by
		\be
			\begin{split}
				\ket{\lambda_\pm}&= \sqrt{\frac{q(r\mp(1-2q))}{r(1\pm r)}} \ket{\psi_0(x_0)} \\
				&\pm \sqrt{\frac{(1-q)(r\pm(1-2q))}{r(1\pm r)}} \ket{\psi_0(x_0+\epsilon)}\, ,
			\end{split}
		\label{eq:Prassad}
		\ee
	where the eigenstates $\{\ket{\lambda_+}, \ket{\lambda_{-}}\}$ correspond to the orthonormal basis $\{\ket{\psi_0(x_c)}, \ket{\psi_0^\perp(x_c)}\}$, where $\ket{\psi_0(x_c)}$ is the fundamental Hermite-Gauss mode centred at the intensity centroid 
	\be
	x_c = x_0 +(1-q)\epsilon.
	\label{eq:Ic}
	\ee  
	
	Observe that both the purity, $r$, and angle, $\theta$, of the Bloch vector ${\bf r}$ are functions of the overlap, $c$, and relative weight $q$.  Hence estimates 
	of $(c,q)$ can be obtained by estimating the purity and/or the Bloch angle $\theta$. The latter is a multi-parameter estimation problem that has been studied 
	extensively in~\cite{Bagan05, Bagan06, de_Vicente10}.  Next we review multi-parameter classical and quantum estimation. 
	
	\subsection{\label{sec:CramerRao}Classical and quantum multi-parameter estimation}

	Let $Y$ be a random variable distributed according to $p(y|\boldsymbol{\theta})$, where $\boldsymbol{\theta}=(\theta_1,\ldots,\theta_M)^T\,\in 
	\Theta\subset \cR^M$ are a set of $M$ parameters.  The goal in parameter estimation is to infer the set of parameters $\boldsymbol{\theta}$ by sampling from 
	the random variable $Y$.  Here, and in what follows, we shall assume each sample is independent and identically distributed.  Let ${\bf y}=(y_1,\ldots,y_N)\in Y^N$ 
	be a realization of the random variable $Y$.  An \emph{estimator} is any function $f: Y^N\to \Theta$ whose value on the realization ${\bf y}$ yields an estimate 
	$\hat{\boldsymbol{\theta}}:=f({\bf y})$ of the parameters $\boldsymbol{\theta}$.  An estimator is said to be \emph{unbiased} if 
		\be
			\int_{Y^N}\, p({\bf y}|\boldsymbol{\theta})\,f({\bf y}) \d{\bf y}= \boldsymbol{\theta},
		\label{eq:unbiased}
		\ee  
	where we have assumed that $Y$ is a continuous random variable~\footnote{In the case of discreet random variables one replaces the integral with a sum.}.
	An estimator is \emph{consistent} if for any $\delta > 0$ and a sequence $f^{(k)}({\bf y})$ of estimates 
		\be
			\lim_{k\to\infty} \mathrm{Pr}\left(\abs{f^{(k)}({\bf y})-\boldsymbol{\theta}}>\delta\right)=0.
		\label{eq:consistency}
		\ee	 
	The \emph{covariance matrix} of an estimator, $\mathrm{Cov}(f({\bf y}))$, with matrix elements given by
		\be
			\mathrm{Cov}_{ij}(f({\bf y}))=\int_{Y^N} p({\bf y}|\boldsymbol{\theta})\, (f_i({\bf y})-\theta_i)(f_j({\bf y})-\theta_j)\, \d{\bf y}\, ,
		\label{eq:Covariance}
		\ee 
	is a measure of an estimators precision.  An estimator is said to be \emph{efficient} if, for any other estimator $g: Y^N\to \Theta$ it holds that 
		\be
			\eta := \mathrm{Cov}(f({\bf y}))\, \mathrm{Cov}^{-1}(g({\bf y})) <\one\, .
		\label{eq:efficiency} 
		\ee	
	For \emph{unbiased} estimators
	a lower bound to the covariance matrix is furnished by a famous theorem by Cram\'er and Rao~\cite{Cramer61}
		\be
    			\mathrm{Cov}(f({\bf y}))\geq  {\bf F}^{-1}\left(p\left({\bf y}|\boldsymbol{\theta}\right)\right),
		\label{eq:CRbound}
		\ee
	where ${\bf F}\left(p\left({\bf y}|\boldsymbol{\theta}\right)\right)$ is the Fisher information matrix~\cite{Fisher22}
		\be
    			F_{ij}(p({\bf y}|\boldsymbol{\theta})):=\hspace{-2mm}\int_{Y^N}  p({\bf y}|\boldsymbol{\theta})\left(\frac{\partial\ln 
			p({\bf y}|\boldsymbol{\theta})}{\partial{\theta_i}}\right) \left(\frac{\partial\ln 
			p({\bf y}|\boldsymbol{\theta})}{\partial{\theta_j}}\right) \d{\bf y}.
    		\label{eq:Classical_Fisher}
		\ee
	 The inequality in \eqnref{eq:CRbound} can always be saturated in the limit of infinite number of samples by the maximum likelihood estimator~\cite{Wilks62}.

	In quantum mechanics the random variable $Y$ describes the outcomes of a quantum mechanical measurement whose probability distribution is given according 
	to Born's rule    
		\be
    			p({\bf y}|\boldsymbol{\theta}):=\tr\left(E_{\bf y}\rho(\boldsymbol{\theta})\right)\, ,
    		\label{eq:Bornrule}
		\ee
	where $\{E_{\bf y}\}$ are any set of positive operator value measures (POVMs) that form a resolution of the identity, $\sum_{\bm{y}} E_{\bm y}=\one$, 
	and $\rho(\boldsymbol{\theta})\in\cB(\cH)$. Under this constraint one can search over the space of all allowable measurements to find those POVMs for 
	which the statistics of the corresponding measurement outcomes achieve the maximum Fisher information.  The latter optimization gives rise to the 
	\emph{Quantum Fisher Information} (QFI) matrix, ${\bf \cF}$, i.e., 
		\be
			\bm{\cF}\left(\rho(\boldsymbol{\theta})\right):=\max_{E_{\bf y}\geq 0, \sum_{\bf y} E_{\bf y}=\one}\, 
			{\bf F}\left(p({\bf y}|\rho(\boldsymbol{\theta})\right),
		\label{eq:QFImatrix}
		\ee 
	and the following set of inequalities regarding the covariance matrix hold 
		\be
			\mathrm{Cov}(f({\bf y}))\geq  {\bf F}^{-1}\left(p\left({\bf y}|\boldsymbol{\theta}\right)\right)\geq  \bm{\cF^{-1}}
			\left(\rho(\boldsymbol{\theta})\right)\, .
  		\label{eq:QuantumCRbound}
		\ee
	The latter inequality is commonly referred as the quantum Cram\'{e}r-Rao bound~\cite{Braunstein94}.  

	An optimal measurement for estimating each of the parameters $\theta_i, \, i\in(1,\ldots, M)$ can be explicitly constructed as follows.  Define the 
	Hermitian operators  $L_{\theta_i}$ satisfying the equation 
		\be
			\frac{\partial\,\rho(\boldsymbol{\theta})}{\partial{\theta_i}}:=\frac{1}{2}\left(L_{\theta_{i}}\rho(\boldsymbol{\theta})+
			\rho(\boldsymbol{\theta})L_{\theta_{i}}\right).
		\label{eq:SLD_def}
		\ee 
	These are the \emph{Symmetric Logarithmic Derivative} (SLD) operators  for each parameter $\theta_i$.  In terms of the eigenbasis 
	$\{\ket{\lambda_k}\}$ of $\rho(\boldsymbol{\theta})\in\cB(\cH)$, the SLD operators take the form
		\be
    			L_{\theta_{i}}=2\sum_{\substack{k,\ell\\ \lambda_k+\lambda_\ell\neq0}}\frac{\bra{\lambda_k}\partial_{\theta_{i}}\rho({\boldsymbol{\theta}})
			\ket{\lambda_{\ell}}}{\lambda_k+\lambda_\ell}\ketbra{\lambda_{k}}{\lambda_{\ell}}.
    		\label{eq:sld}
		\ee
	In turn the QFI matrix elements read
		\be
     			\bm{\cF}_{ij}\left(\rho(\boldsymbol{\theta})\right)= \mathrm{Re}\left(\tr\left(L_{\theta_{i}}L_{\theta_{j}}\rho(\boldsymbol{\theta})\right)\right). 
     		\label{eq:QFIij}
		\ee
	Observe that, in general, the SLD operators need not commute, in which case the quantum Cram\'er-Rao bound is not asymptotically achievable. A necessary and 
	sufficient condition for asymptotically saturating the quantum Cram\'er-Rao bound is~\cite{Ragy16} 
		\be
			\tr\left(\rho(\boldsymbol{\theta})[L_{\theta_i},L_{\theta_j}]\right)=0.  
		\label{eq:saturability}
		\ee
	
	Whilst the parameters $\boldsymbol{\theta}$ are directly accessible via measurement, often times what we are interested in is to estimate some function 
	${\bf h}:\Theta\to \cR^M$ of the parameters.  Given the covariance matrix, $\mathrm{Cov}(f({\bf y}))$, one can obtain the covariance 
	matrix, $\mathrm{Cov}({\bf h(\boldsymbol{\theta})})$, by standard error propagation.  Specifically, the two covariance matrices are related by
		\be
			\mathrm{Cov}({\bf h(\boldsymbol{\theta})})=\cJ\, \mathrm{Cov}(f({\bf y}))\, \cJ^{T}, 
		\label{eq:error_propagation}
		\ee 
	where $\cJ$ is the $M\times M$ Jacobian matrix whose elements are given by 
		\be
			\cJ_{kl}=\frac{\partial h_k(\boldsymbol{\theta})}{\partial\theta_l}\, .
		\label{eq:Jacobian}
		\ee	
		
\section{\label{sec:QFI} QFI for purity and relative weight estimation in a single boson}
	
	We are interested in estimating the purity and relative intensity, i.e.,  $\boldsymbol{\alpha}=(r,q)^T$, of the Bloch vector given by \eqnsref{eq:Blochvector}
	{eq:angle}. The corresponding SLD operators can be written as
		\begin{align}\nonumber
			L_r&=a_0\, \one + {\bf a}\cdot\paulivec\\
			L_q&=b_0\, \one + {\bf b}\cdot\paulivec\,,
		\label{eq:SLDbloch} 
		\end{align}	
	where $a_0,b_0\in \R$, and ${\bf a},\, {\bf b}\in\R^3$.  Substituting \eqnref{eq:SLDbloch} in \eqnref{eq:SLD_def} one arrives at the following set of 
	equations 
		\begin{align}\nonumber
			a_0 &= -{\bf a}\cdot{\bf r}\; &\frac{\partial {\bf r}}{\partial r}= a_0{\bf r}+{\bf a}\\
			b_0 &= -{\bf b}\cdot{\bf r}\; &\frac{\partial {\bf r}}{\partial q}= b_0{\bf r}+{\bf b}\, .
		\label{eq:SLDeqns}
		\end{align}
	The solutions to \eqnref{eq:SLDeqns} are 
		\begin{align}\nonumber
			a_0&=-\frac{r}{1-r^2}\; &{\bf a}=\frac{1}{1-r^2}\, \hat{{\bf r}} \\
			b_0&= \frac{1-2q}{2q(1-q)}\; &{\bf b}=\bematrix -\frac{c\sqrt{1-c^2}}{q}\\ 0\\ \frac{1-2c^2(1-q)}{2q(1-q)}\ematrix\, ,
		\label{eq:SLDvecs}
		\end{align}	
	and the corresponding QFI matrix reads
		\be
			\begin{split}
				\bm{\cF}(\rho)&=\bematrix
 					    a^2-a_0^2 & {\bf a}\cdot{\bf b}-a_0 b_0 \\
					    {\bf a}\cdot{\bf b}-a_0 b_0 & b^2-b_0^2
					    \ematrix\\
					    & = \bematrix \frac{1}{1-r^2} & \frac{2q-1}{2q(1-q)r}\\ \frac{2q-1}{2q(1-q)r} & \frac{1-c^2}{q(1-q)}\ematrix\, ,
			\end{split}
		\label{eq:QFI}
		\ee
	where $a^2={\bf a}\cdot{\bf a}$ and similarly for $b^2$. 
		
	To obtain the QFI matrix corresponding to the estimation of the separation $\epsilon$ and relative intensity $q$, we note that in the regime
	$\epsilon\ll\ 1$ the overlap can be expressed as $c^2= 1-\frac{\epsilon^2}{4}$. Substituting the latter into \eqnref{eq:purity} the functional dependence of the 
	parameters we wish to estimate, $(\epsilon, q)$, to the parameters measured are
        \begin{align}
            h_1(r, q)&:=\epsilon= \sqrt{\frac{1-r^2}{q(1-q)}}\\
            h_2(r, q)&:= q = q \, .
        \end{align}
        Hence the elements of the Jacobian matrix are given by 
		\begin{align} \nonumber
			\frac{\partial \epsilon}{\partial r}&=-\frac{r\epsilon}{1-r^2}\quad  &\frac{\partial\epsilon}{\partial q}&=-\frac{(1-2q)\epsilon}{2q(1-q)}\\
			\frac{\partial q}{\partial r}&= 0 \quad     &\frac{\partial q}{\partial q}&=1 \,.
		\label{eq:derivatives}
		\end{align}
	The  QFI matrix for ${\bm h}(\boldsymbol{\alpha})$ reads, to leading order in $\epsilon$
		\be
			\bm{\cF}({\bm h}(\boldsymbol{\alpha}))=\bematrix (1-q) q+\cO(\epsilon^{\frac{3}{2}}) & \frac{(1-2 q)}{2} \epsilon +\cO(\epsilon^{\frac{5}{2}})  \\
 							\frac{(1-2 q)}{2} \epsilon +\cO(\epsilon^{\frac{5}{2}})  & \frac{\epsilon^2}{16q(1-q)}  \ematrix\, .
		\label{eq:QFI_epsilon}												
		\ee
	Observe that $\bm{\cF}_{00}({\bm h}(\boldsymbol{\alpha})) = (1-q) q$---corresponding to the QFI for the separation $\epsilon$ of the signals---is constant and 
	independent of $\epsilon$.  Thus the Rayleigh limit is entirely circumvented.  Note that the measurements furnishing optimal estimation for purity and relative 
	intensity are incompatible since as 
		\be
			[L_r,L_q]=-4\ii\left(\frac{c\sqrt{1-c^2}}{r(1-r^2)}\right)\, \sy\,.
		\label{eq:incompatibility}
		\ee       
	If instead of estimating $\boldsymbol{\alpha}=(r,q)^T$, we try to infer the separation 
	and relative intensity of the sources by estimating $\boldsymbol{\alpha}=(r,\theta)^T$ we obtain a suboptimal estimate for the relative intensity $q$ (see 
	\appref{sec:Len}).	
       
	Hitherto our analysis assumed that the location $x_0$ for one of the sources is known and that our demultiplexing 
   	measurement can be placed perfectly at the intensity centroid $x_C$.  We now relax this condition to the more realistic situation where the location parameter is 
	not known perfectly and/or the demultiplexing measurement cannot be perfectly placed at this position.  To that end, let 
	$\{\ket{\psi_0(y)},\ket{\psi_0^\perp(y)} \}$ be the orthonormal basis corresponding to some position $y\in\R$.  For $y\in (x_0\pm\epsilon)$ there 
	corresponds a unitary operator $U(\alpha):=e^{-\ii\frac{\alpha}{2}\sy}$ where $\alpha \in(-\frac{\pi}{2}, \, \frac{\pi}{2})$ is the polar angle connecting the 
	direction defined by the basis $\{\ket{\psi_0(y)},\, \ket{\psi_0^\perp(y)}\}$ to that defined by the basis $\{\ket{\psi_0(x_0)},\,\ket{\psi_0^\perp(x_0)}\}$.  
	If $y\in (x_0\pm\epsilon)$ is known then the description of the state $\rho$ in \eqnref{eq:finite_mixture} is given by
		\be
			\rho' = U(\alpha)\, \rho\, U(\alpha)^\dagger\, .
		\label{eq:rotated_rho}
		\ee
	If, however, the amount of misalignment is not known then the proper description of the state of the two signals is given by 
		\be
			\begin{split}
				&\rho_{\mathrm{avg}} = \frac{1}{2}\int_{-\frac{\pi}{2}}^{\frac{\pi}{2}}\,\cos(\alpha)\, U(\alpha)\, \rho\,U(\alpha)^\dagger\, \d\alpha\\
								    & = \frac{1}{2}\left(\one + {\bf r}_{\mathrm{avg}}\cdot\paulivec\right)	\, , 
			\end{split}
		\label{eq:avg_state}
		\ee
	where $\frac{1}{2}\cos(\alpha)\d\alpha$ is the uniform distribution over the polar angle, ${\bf r}_{\mathrm{avg}}=\frac{\pi}{4}\, {\bf r}$, and we have chosen, 
	without loss of generality, the $\hat{\bf z}$-direction to correspond to the basis associated with the reference position $x_0$.
       
       A similar computation for the average state $\rho_{\mathrm{avg}}$ of \eqnref{eq:avg_state} yields the following solution for the SLD operators
		\begin{align} \nonumber
			& a^{(\mathrm{avg})}_0=-\frac{r_{\mathrm{avg}}}{1-r_{\mathrm{avg}}^2}, \; b^{(\mathrm{avg})}_0= \frac{\pi^2}{8}\frac{(1-2q)(1-c^2)}
			{1-r_{\mathrm{avg}}^2}\\ 
			\nonumber
			&{\bf a}_{\mathrm{avg}}=\frac{1}{1-r_{\mathrm{avg}}^2}\, \hat{{\bf r}}_{\mathrm{avg}}\\
			&{\bf b}_{\mathrm{avg}}=\frac{\pi\sqrt{1-c^2}}{2(1-r_{\mathrm{avg}}^2)}\bematrix c\left(1+\frac{\pi^2}{16}(1-2q-2c^2(1-q))\right)\\ 0\\ 
			\sqrt{1-c^2}(1-\frac{\pi^2}{8}(1-q)c^2)\ematrix\, .
		\label{eq:SLDvecavg}
		\end{align}
	 The QFI matrix for estimating the average purity and relative intensity reads 
		\be
			\bm{\cF}(\rho_{\mathrm{avg}})=\bematrix
 						\frac{1}{1-r_{\mathrm{avg}}^2} & -\frac{\pi^2}{8}\frac{(1-2q)(1-c^2)}{r_{\mathrm{avg}}(1-r_{\mathrm{avg}}^2)}  \\
 						-\frac{\pi^2}{8}\frac{(1-2q)(1-c^2)}{r_{\mathrm{avg}}(1-r_{\mathrm{avg}}^2)} & \frac{\pi^2}{4}\frac{(1-c^2)(1-\frac{\pi^2}{16}c^2)}
						{r_{\mathrm{avg}}^2} \ematrix\, ,
		\label{eq:QFI_avg1}												
		\ee
		this result is exact, depending on the overlap $c$, the relative intensity $q$ and the average purity $r_{\mathrm{avg}}$. Including the overlap dependence on the separation, the QFI matrix for estimating the separation and relative intensity up to leading order in $\epsilon$ is
	%whilst the QFI matrix for estimating the separation and relative intensity becomes, to leading order in $\epsilon$,
		\be
			\bm{\cF}({\bm h}(\boldsymbol{\alpha}))= \bematrix
 			q(1-q) & \frac{\pi}{2}(1-2 q)\sqrt{\frac{q(1-q)}{\pi^2-16}}\epsilon^2 \\
 			\frac{\pi}{2}(1-2 q)\sqrt{\frac{q(1-q)}{\pi^2-16}}\epsilon^2 & \frac{\pi ^2 \epsilon ^2}{16} \ematrix\, .
		\label{eq:QFI_avg2}
		\ee
	Once again, observe that $\bm{\cF}_{00}({\bm h}(\boldsymbol{\alpha})) = (1-q) q$, meaning that super-resolution for separation estimation is again restored.  
	However, under misalignment the estimation of the relative intensity $q$ is suboptimal and does not even depend on $q$. We note that \eqnsref{eq:QFI_avg1}
	{eq:QFI_avg2} correspond to the best possible precision under misalignment when the bosons are measured one at a time.
	Just as in the case where the location of the sources is known, the measurements for estimating the purity and relative weight of the signals are incompatible. 
	However, the necessary and sufficient condition, $\tr\left([L_r,L_q]\rho\right)=0$, for achieving the quantum Cram\'{er}-Rao bound holds,  which opens the possibility for a collective measurement strategy.

\section{\label{sec:collective}Collective, super-resolving quantum measurements}

	We now introduce a collective measurement strategy that asymptotically saturates the multi-parameter quantum Cram\'er-Rao bound for separation and relative 
	intensity estimation.  The measurement in question is the Schur-sampling measurement---the projections onto the total angular momentum operator $\bf J^2$ 
	followed by the projection along some direction, $\hat{\bf n}$. This measurement is ubiquitous in quantum estimation and 
	information theory~\cite{Alicki88,Keyl01,Buhrman01,Ekert02,Bagan05,Cincio18,Badescu19,Fanizza20}, and for the case of $N=2$ bosons is equivalent to the 
	Hong-Ou-Mandel effect~\cite{GarciaEscartin13}.  In \appref{sec:spectrum}, we provide a review of this measurement strategy, including an example of its 
	implementation for super resolution using the Hong-Ou-Mandel effect for the case of $N=2$ bosons.
		
	Using the invariance under permutations, $N$ copies of the state $\rho$ in \eqnref{eq:Bloch_form} can be written as 
		\be
			\rho^{\otimes N}=\bigoplus_J p(J)\, \omega^{(J)}\otimes \frac{\one}{d_J},
		\label{eq:block_diagonal}
		\ee
	where the probability distribution $p(J)$, as well as the normalized states $\omega(J)$ and constant $d_J$, are given in \eqnsref{eq:pJtauJ}{eq:dimension_J} of 
	\appref{sec:spectrum}.  The collective measurement we shall perform consists of the following projectors 
	$\{\Pi_{J,M}=\ket{J,M}_{\hat{\bf n}}\bra{J,M}\, \vert \, 0(\frac{1}{2})\leq J\leq \frac{N}{2}, \, -J\leq M\leq J\}$, where $J$ is the total angular momentum of 
	$N$ qubits, and $M$ is the component of the angular momentum along some direction $\hat{\bf n}$. The joint probability distribution of obtaining the 
	measurement outcome $J, M$ is given by
		\be
			p(J,M|r,q) = p(J|r) \, p(M|J,q)
		\label{eq:joint_prob}
		\ee 	
	with $p(J|r)$ given in \eqnref{eq:pJtauJ}. 
	
	We pause here briefly to clarify the form of \eqnref{eq:joint_prob} and in particular its dependence on  $r$ and $q$.  The marginal distribution 
	$p(J|r)=\sum_{M=-J}^J p(J,M\vert r,q)$ depends solely on the purity $r$, of $\rho$.  Indeed as was shown in~\cite{Bagan05} in the limit of asymptotically large 
	number, $N$, of systems this probability distribution furnishes an asymptotically unbiased estimator of the purity whose Fisher information approaches the QFI 
	value of $(1-r^2)^{-1}$.  This estimate, along with \eqnref{eq:purity} is then used to express any subsequent dependence on $r$ in $p(M|J,q)$ solely in terms of 
	$q$.  Notice, however, that because of \eqnref{eq:purity} any uncertainty in $q$ also affects the uncertainty of $r$.  
	
	Hence the matrix elements of the Fisher information matrix for the collective measurement $\{\ket{J,M}_{\hat{n}}\}$ reads
		\be
			\begin{split}
				F_{rr} &= \sum_{J=0(\nicefrac{1}{2})}^{\frac{N}{2}}\, \frac{\left(\frac{\d p(J|r)}{\d r}\right)^2}{p(J|r)}\\
				F_{qq} & = \sum_{J=0(\nicefrac{1}{2})}^{\frac{N}{2}}\sum_{M=-J}^J \, \frac{\left(\frac{\partial p(J, M|r, q)}{\partial q}\right)^2}{p(J, M|r, q)} \\
				F_{rq} & = \sum_{J=0(\nicefrac{1}{2})}^{\frac{N}{2}}\sum_{M=-J}^J\, \frac{\frac{\d p(J|r)}{\d r}\frac{\partial p(J,M|r,q)}{\partial q}p(M|J,q)}{p(J, 
				M|r, q)}\, .
			\end{split}
		\label{eq:Fishercollective measurement}
		\ee
	As ${\bf F} = \sum_{J} {\bf F}^{(J)}$ it follows that to maximize the total Fisher information, one can maximize the Fisher information in every sector separately.  
	Hence, the direction ${\hat{n}}$ in which one measures in order to obtain information about the relative intensity $q$ can depend on the value of $J$.	
	Here, we analyze the statistics of the Schur-sampling measurement and demonstrate that it saturates the quantum Cram\'er-Rao bound---both for known 
	(\secref{sec:Sxc}) and unknown (\secref{sec:Save}) intensity centroid (\eqnref{eq:Ic})---when estimating the separation and relative intensity of $\rho$ 
	in \eqnref{eq:finite_mixture}, as the number of collectively measured bosons increases.
	
	\subsection{\label{sec:Sxc} Spectrum measurement with known intensity centroid}
	
	To determine the probability distribution $p(J,M|r,q)$ we choose, without loss of generality to associate the orthonormal basis  
	$\{\ket{\psi_0(x_0)},\, \ket{\psi_0^\perp(x_0)}\}$ with the eigenbasis of $\sz$, i.e., $\ket{\psi_0(x_0)}=\ket{\frac{1}{2},\frac{1}{2}}_{\hat{\bf z}}, \, 
	\ket{\psi_0^\perp(x_0)} = \ket{\frac{1}{2},-\frac{1}{2}}_{\hat{\bf z}}$.  The eigenstates of $\rho$ (\eqnref{eq:eigenstates}) are then given by 
		\be
			\ket{\lambda_\pm}=\ket{\frac{1}{2},\pm\frac{1}{2}}_{\hat{\bf r}} = \sum_{m=-\frac{1}{2}}^{\frac{1}{2}}\, D^{(\frac{1}{2})}_{\pm\frac{1}{2},m}(\theta)
			\ket{\frac{1}{2},m}_{\hat{\bf z}}
		\label{eq:eigenstates_rho}
		\ee
	where $D^{(\frac{1}{2})}(\theta)$ are the Wigner small-$d$ matrices, an irrep of $\mathrm{SU(2)}$ corresponding to the irrep label $J$,  and $\theta$ is given 
	by \eqnref{eq:angle}.  Consequently the eigenstates $\ket{J,M}_{\hat{\bf r}}$ of $\omega^{(J)}$ in \eqnref{eq:block_diagonal} are given by 
		\be
			\ket{J,M}_{\hat{\bf r}}=\sum_{K=-J}^J D^{(J)}_{K, M}(\theta)\, \ket{J,K}_{\hat{\bf z}}\, 
		\label{eq:AngMombasis}
		\ee
	where $\ket{J,M}_{\hat{\bf z}}$ are the eigenstates of the operator $J_{\hat{\bf z}}$. Similarly, we may write the total angular momentum measurement and its 
	projection along some direction $\hat{\bm b}$ defined in \eqnref{eq:SLDvecs} as  
		\be
			\ket{J,M}_{\hat{\bm b}}=\sum_{K=-J}^J D^{(J)}_{K, M}(\phi)\, \ket{J,K}_{\hat{\bf z}}\, ,
		\label{eq:AngMommeas}
		\ee
	where 
		\be
			\phi = -\cos^{-1}\left(\frac{1-2c^2(1-q)}{\sqrt{1-4q(1-q)c^2}}\right)\, .
		\label{eq:phi}
		\ee
	
	It follows that the probability defined in \eqnref{eq:joint_prob} in this scenario is
		\be
			p(M|J,q) = \frac{1}{Z_J}\sum_{K=-J}^{J}\, R^K \, D^{(J)}_{K,M}\left(\theta-\phi \right)^2\, ,
		\label{eq:probWigner}
		\ee	
		where $Z_J$ and $R^K$ are defined in  \eqnref{eq:pJtauJ}.
		
\begin{figure}[ht!]  
   			\centering
   				\includegraphics[width=0.48\textwidth]{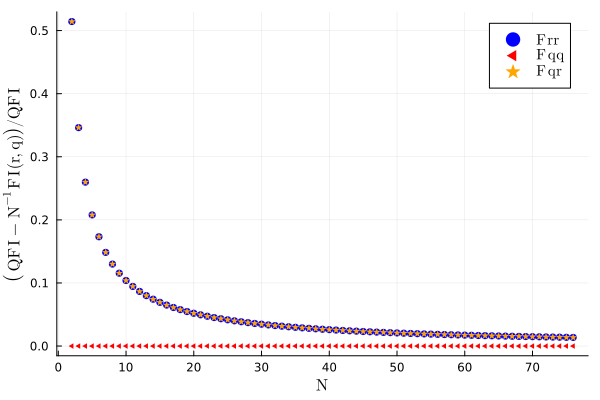}
   			\caption{Relative error difference between the matrix elements of the Fisher information for the measurement in \eqnref{eq:AngMommeas} and the QFI 
			of \eqnref{eq:QFI_epsilon} as a function of the number of copies. The overlap and relative intensity are set to  $c = 0.97$ and $q = 0.8$ respectively.}  
   			\label{fig:Fallb}
		\end{figure}

	In \figref{fig:Fallb} we plot the ratio $\frac{\cF_{ij}-N^{-1} F_{ij}}{\cF_{ij}}, \, i,j\in\{r,q\}$ for several values of $2\leq N\leq 76$ for a measurement of the 
	total angular momentum and its projection along $\hat{\bm b}$ (the optimal direction for estimating $q$).  Observe that $N^{-1}F_{qq}=\cF_{qq}$ for all values of 
	$N$, whereas the Fisher information for $r$ and the correlations $F_{rq}$ approach those of the optimal QFI as more and more bosons are measured collectively.  
	Hence, this measurement asymptotically saturates the QFI, allowing us to simultaneously estimate both the separation and relative intensity of the point sources, 
	something that is impossible to achieve processing the bosons one at a time.  
	A more detailed explanation on the implementation is given in  \secref{sec:Implementation}.

	\subsection{\label{sec:Save} Spectrum measurement with unknown intensity centroid}
	
	Let us now consider performing a collective measurement of the $N$ bosons in the presence of misalignment.  The collective state 
	is now given by 
		  \be
				\rho'^{\otimes N}=\bigoplus_J p(J)\, \int_{-\frac{\pi}{2}}^{\frac{\pi}{2}}\, \cos(\alpha)\, U^{(J)}(\alpha)\,\omega^{(J)}\, U^{(J)\dagger}(\alpha)
				\otimes \frac{\one}{d_J},
		\label{eq:block_diagonal_avg}
		\ee
	where we have made use of the fact that $U^{\otimes N}(\alpha)\cong \oplus_J U^{(J)}(\alpha)$.  The collective spectrum measurement is given by
		\be
			\ket{J,M}_{\hat{\bm b}}=\sum_{K=-J}^J D^{(J)}_{K, M}(\phi)\, \ket{J,K}_{\hat{\bf z}}\, .
		\label{eq:AngMommeasavg}
		\ee 
	resulting in the probability distribution   
		\be
			p(M|J,q) = \frac{1}{Z_J}\sum_{K=-J}^{J}\, R^K \left[\frac{1}{2}\int \d\alpha \,\mathrm{cos}(\alpha)\,D^{(J)}_{K,M}\left(\theta-\phi+
			\alpha\right)^2\right]\, ,
		\label{eq:probWigner2}
		\ee	
	\normalsize
	with the marginal distribution $p(J)$---which allows for an estimate of the purity---being the same as before.  Observe that we are free to optimize over the 
	angle $\phi$---or equivalently the direction $\hat{\bm b}$---for each sector $J$ separately in order to maximize the Fisher information and indeed we find that 
	the optimal direction to measure in this case is different depending on the value of $J$ obtained.  Never the less, choosing 
		\be
  			\phi = \phi_{\mathrm{avg}}=-\mathrm{cos}^{-1}\left( \frac{2\sqrt{(1 -  c^2)}(8-(c\pi)^2(1 - q))}{\sqrt{256 - (c\pi)^2(32 + r^{2}_{avg})}}\right).
		\label{eq:phiavg}
		\ee	
	for all values of $J$ yields a Fisher information for $F_{qq}$ and $F_{rq}$ that differ from the optimal ones in each sector $J$ only in the third decimal place.
	
\begin{figure}[ht!]  
   			\centering
   				\includegraphics[width=0.48\textwidth]{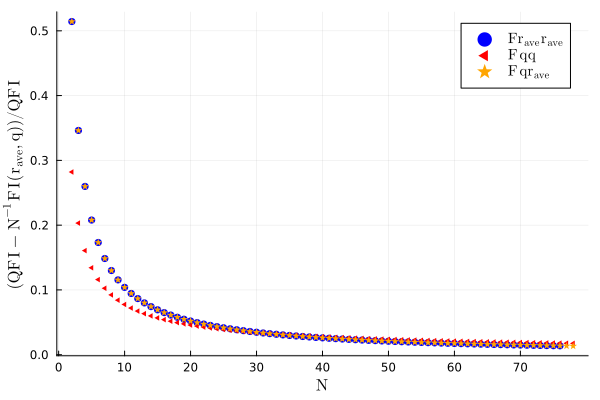}
   			\caption{Relative error between the matrix elements of the Fisher information for the measurement in \eqnref{eq:AngMommeasavg} and the QFI of 
			\eqnref{eq:QFI_epsilon} as a function of the number of copies. The overlap and relative intensity are set to  $c = 0.97$ and $q = 0.8$ respectively.}  
   			\label{fig:Fallbave}
		\end{figure}

	In \figref{fig:Fallbave} we plot the numerically evaluated $\frac{\cF_{ij}-N^{-1} F_{ij}}{\cF_{ij}}, \, i,j\in\{r,q\}$ for all $2\leq N\leq 76$ for a spectrum 
	measurement with angle $\phi_{\mathrm{avg}}$ defined in \eqnref{eq:phiavg} for the case of unknown misalignment. Observe that both $F_{qq}$
	as well as $F_{rq}$ asymptotically approach $\cF_{qq},$ and $\cF_{rq}$ respectively. 
	Our result shows the power of the spectrum measurement, as $\bf J^2$ relies solely on symmetry under particle exchange, this measurement is inherently robust 
	against misalignment in providing an estimate of the purity. Furthermore, the relative intensity measurement demonstrates how the intensity centroid dependence 
	can be circumvented and still approach the QFI asymptotically something which is impossible to do with misaligned measurements that process bosons one at a time 
	(see \eqnref{eq:QFI_avg2}). In the next section we discuss possible implementations of the collective measurement its advantages and drawbacks.

\section{\label{sec:Implementation}Implementations of collective measurement schemes}
	
\begin{figure}[ht!]  
   \centering
   	\includegraphics[width=0.48\textwidth]{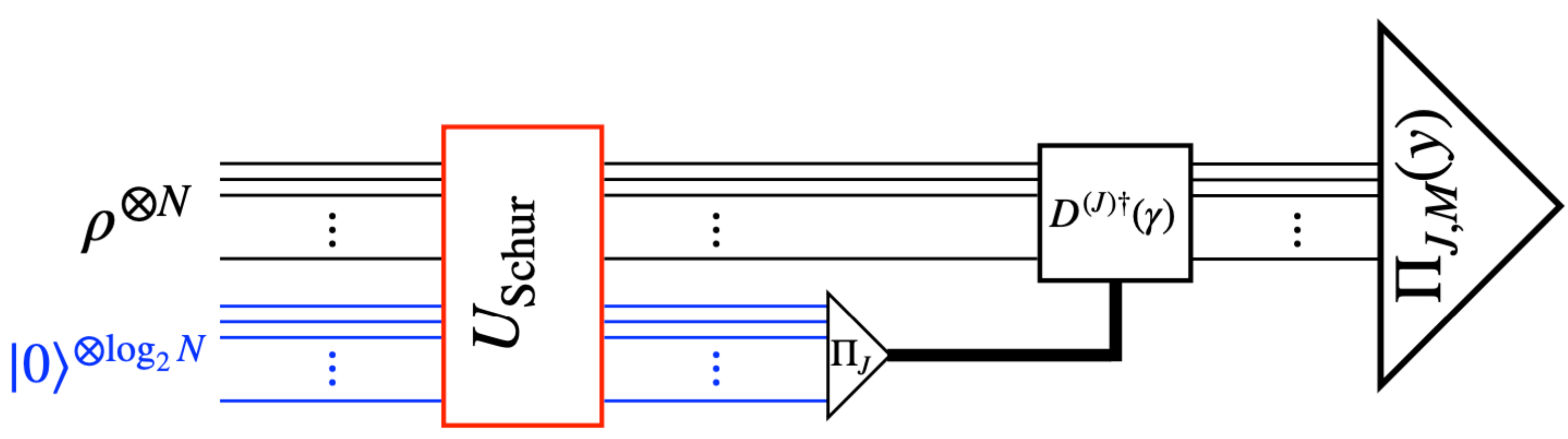}
   	\caption{A circuit representation of the protocol implementing our super resolving collective measurement strategy for estimating the purity and relative intensity 
	of two incoherent bosonic sources. The Schur transform $U_{\mathrm{Schur}}$ is applied to $N$ bosons each of which in the state $\rho$ of 
	\eqnref{eq:finite_mixture} (black lines) and $\log_2(N)$ qubits initialized in the state $\ket{0}$ (blue lines). The Schur transform imprints the value of the total 
	angular momentum $J$ in binary form on the ancilla qubits.  Upon measuring the latter (small triangle), the classical value $J$ is used to implement the required 
	symmetrized $\mathrm{SU}(2)$ rotation $D^{(J)\dagger}(\gamma)$ (\eqnref{eq:AngMommeas}), where for a known intensity centroid $\gamma=\theta-\phi$ 
	after which the bosons are measured in the $\{\ket{\psi_0(x_c)}, \ket{\psi_0^\perp(x_c)}\}$ basis, whereas for the arbitrary intensity centroid $
	\gamma=\theta-\phi_{\mathrm{avg}}$ followed by a Haar-measurement in different values of $y$ $\{\ket{\psi_0(y)}, \ket{\psi_0^\perp(y)}\}$.}
   	\label{fig:Schur}
\end{figure}

	We now outline possible experimental implementations of our proposed measurement scheme for super resolving collective measurements.  Our proposal consists of 
	first transferring the collective state of $N$ bosons to a quantum memory, pre-processing their relevant degrees of freedom before performing the requisite
	Schur-sampling measurement (see~\figref{fig:Schur}). 
	
	The state $\rho^{\otimes N}$ does not live in the bosonic symmetric subspace. It describes the semi-classical state of $N$ bosons multiplexed in different modes 
	(spatial, temporal, polarization or frequency). Depending on which mode the state is multiplexed, different quantum memories are optimal for the task. For example, 
	spatial, temporal, and polarization modes can be stored faithfully using electron
	spins~\cite{Kroutvar2004,Simon2007,Sangouard2011,Sinclair14,Yang18,Lago2021,Bland-Hawthorn21} or atomic 
	ensembles~\cite{Julsgaard2004,Fleischhauer2002,Chaneliere2005,Mazelanik22}, whereas frequency modes can be mapped into NV centers~\cite{Maurer2012} or 
	trapped ions \cite{Kielpinski01,Biercuk2009}.

	After the storage of the collective state of $N$ bosons, the Schur transform must be implemented. The latter is known to be efficiently implementable up to an 
	error $\varepsilon>0$ using $O(\mathrm{poly}(N, \ln d, \ln \frac 1 \varepsilon))$ single and two-qubit gates and requires an additional $\mathcal{O}(\ln N)$ 
	qubit auxilliary systems~\cite{Bacon07}.  The resulting state of the $N$ bosons after the Schur transform (\eqnref{eq:finite_mixtureblock}) can be written as  
	\be
		U_{\mathrm{Schur}} \,\rho^{\otimes N}\, U^\dagger_{\mathrm{Schur}}=\sum_{J} p(J)\, \proj{J}\otimes \omega^{(J)}\otimes\frac{\one}{\abs{\cP_J}}\,,
		\label{eq:practicalSchur} 
	\ee
	where the $N$ dimensional register $\{\ket{J}\}$ stores the value of the random variable $J$ which is distributed according to $p(J)$ and gives an estimate of 
	the purity.  
	
	In order to estimate $q$, for a known intensity centroid we use the result of estimating $\{\ket{J}\}$ to control the unitary operator 
	$D^{(J)\dagger}(\theta-\phi)$. This unitary corresponds to a rotation by an angle $\theta-\phi$ in the angular momentum basis (as defined in \eqnsref{eq:angle}
	{eq:phi}), which effectively alters the relative intensity between different $J$ modes. Finally, if the intensity centroid is known each boson is demultiplexed into 
	the Hermite-Gauss modes $\{\ket{\psi_0(x_c)},\ket{\psi_0^\perp(x_c)}\}$, this is equivalent to a B-SPADE measurement, yielding the QFI in 
	\eqnref{eq:QFI_epsilon}. 
	
	A similar procedure can be applied for the case of unknown misalignment. In this case there are two ways of implementing the random Haar-measurement: the first 
	involves after the controlled unitary operator $D^{(J)\dagger}(\theta-\phi_{\mathrm{avg}})$,  to demultiplex each boson into Hermite-Gaussian modes located at 
	different positions $y$, in measurements in the basis $\{\ket{\psi_0(y)},\ket{\psi_0^\perp(y)}\}$ spanned by $\ket{\psi_0(x_0)},\ket{\psi_0(x_0+\epsilon)}$, 
	with their respective weights as in~\eqnref{eq:avg_state} (where $\alpha$ corresponds to $y$). The second method involves implementing the controlled unitary 
	rotations $D^{(J)\dagger}(\theta-\phi_{\mathrm{avg}}+\alpha)$ for different  $\alpha$ as in~\eqnref{eq:avg_state} and projecting in the basis 
	$\{\ket{\psi_0(y)},\ket{\psi_0^\perp(y)}\}$ for a fixed value of $y$. 
	
	A crucial component in the implementation of the Schur-sampling measurement for large number of boson $N$ is storing the all bosons in a quantum memory. 
	Efficient implementations for storing optical modes of light are atomic spin ensembles~\cite{Julsgaard2004,Fleischhauer2002,Chaneliere2005} which have already 
	been used to store 
	displaced gaussian states in the demonstration of imaging~\cite{Parniak18} and spectroscopic~\cite{Mazelanik22} super resolution. Each photon's state can be 
	efficiently transferred onto the spin-state of the atomic ensemble in a three-step process involving first an interaction of the signal photons and an 
	entangling field which entangles the spin ensemble with the photons. Measurement of the signal photons and suitable feedback operations on the atomic ensemble 
	completes the state transfer of the photonic state onto the collective spin state of the memory. The Schur-sampling measurement now corresponds to measuring 
	the collective spin state of the spin ensemble. Such measurements form part of the standard toolkit of spin polarization
	spectroscopy~\cite{Koschorreck10,Colangelo17}.  The post-measurement spin state $\omega^{(J)}$---which recall is a product state---can then be retrieved and 
	coupled into a spatial mode demultiplexer implementing the necessary Hermite-Gauss mode measurement for estimating the relative intensity $q$. Alternatively, 
	for $N<5$ the Schur-sampling of optical modes, the spectral decomposition can be implemented with multi-core fibers~\cite{Carine20}, followed by a post-
	processing of the demultiplexed modes, modulating their relative intensity and a projective measurement in the spatial degree of freedom or a Haar measurement in 
	different centroid positions.

\section{\label{sec:Conclusions} Conclusions}
	
	In this work we have constructed a collective measurement strategy that saturates the multi-parameter Cram\'er-Rao bound for simultaneously estimating the 
	separation and relative intensity of two incoherent point sources. Our measurement leverages the permutation symmetry of the collective state of $N$ bosons to 
	achieve super-resolution in estimating the separation between two incoherent signals without requiring knowledge of the intensity centroid. For signals of equal 
	intensity, the state’s permutation symmetry alone is sufficient to achieve super-resolution in estimating the sources separation. For incoherent sources with unequal 
	intensities, we have shown that additional coarse graining of the measurement enables optimal estimation of the relative intensities. Furthermore, we propose viable 
	implementations of these collective measurements using atomic spin-ensemble quantum memories, multi-port optical devices, and linear optical elements.
	
	Interesting questions for the future are whether such collective measurements can be of use in the case of partially coherent point
	sources~\cite{Hradil21,Wadood21}, or to optical imaging in the near field. It is also interesting to investigate the performance of such collective measurements in 
	the case where the systems under question are fermions as opposed to bosons.  On the more practical side of things, it is crucial to analyse the 
	performance of such collective measurement strategies under realistic experimental conditions, such as coupling losses between the atomic recall stage and the 
	multi-port optical devices, noise and decoherence in the atomic memory implementation and detector imperfections.  These subjects will be addressed in
future work.
\section{Acknowledgements}
	
	We thank M. Parniak, M. Mitchell, A. Lvovsky, and J.~A.~Slater for their insights with regards to the experimental implementation of our collective measurement scheme, particularly on the part of quantum memories.  JOA acknowledges support from EU HORIZON-MSCA-2024-PF-01 project No 101205617 - GATOR, JOA and ML acknowledges support from: European Research Council AdG NOQIA; MCIN/AEI (PGC2018-0910.13039/501100011033, CEX2019-000910-S/10.13039/501100011033, Plan National STAMEENA PID2022-139099NB, project funded MCIN and  by the “European Union NextGenerationEU/PRTR" (PRTR-C17.I1), FPI); QUANTERA DYNAMITE PCI2022-132919, QuantERA II Programme co-funded by European Union’s Horizon 2020 program under Grant Agreement No 101017733; Ministry for Digital Transformation and of Civil Service of the Spanish Government through the QUANTUM ENIA project call - Quantum Spain project, and by the European Union through the Recovery, Transformation and Resilience Plan - NextGenerationEU within the framework of the Digital Spain 2026 Agenda; MICIU/AEI/10.13039/501100011033 and EU (PCI2025-163167); Fundaci\'{o} Cellex; Fundaci\'{o} Mir-Puig; Generalitat de Catalunya (European Social Fund FEDER and CERCA program; Barcelona Supercomputing Center MareNostrum (FI-2023-3-0024); Funded by the European Union (HORIZON-CL4-2022-QUANTUM-02-SGA, PASQuanS2.1, 101113690, EU Horizon 2020 FET-OPEN OPTOlogic, Grant No 899794, QU-ATTO, 101168628),  EU Horizon Europe Program (No 101080086 NeQSTGrant Agreement 101080086 — NeQST)

\bibliography{superres}

\appendix

\section{\label{sec:Len}Estimating purity and relative angle}

  The measurement of estimating the purity $r$, and angle $\theta$ of the state $\rho$ has been shown to be optimal for simultaneously estimating the separation and centroid of two equally bright bosonic 
  sources~\cite{Len22}. In this case the SLD operator for purity is the one derived in \eqnsref{eq:SLDbloch}
    	{eq:SLDeqns} whilst the SLD operator for the estimating $\theta$ is given by 
        \begin{equation}
                L_\theta = s_0 \one +\bm{s}\cdot\bm{\sigma}
        \end{equation}
  with $s_0= 0$ and $\bm{s} = r\, \hat{\bm{r}}_\perp$.  The QFI matrix for estimating the parameters $(r,\theta)$ explicitly reads
        \begin{equation}
            \bm{\cF}(\rho) = \bematrix  \frac{1}{1-r^2} & 0\\ 0 & r^2 \ematrix\, .
        \end{equation}
  Note that just as in the case of purity and relative intensity estimation the SLD operators $L_r, L_\theta$ do not commute and hence cannot be measured simultaneously.

  To obtain the QFI matrix corresponding to the estimation of the separation $\epsilon$ and relative intensity $q$, we again consider the limit of highly overlapping signals, $\epsilon\ll\ 1$, 
  and substitute into \eqnsref{eq:purity}{eq:angle}. The purity $r$ and Bloch angle $\theta$ are related to $\epsilon$ and $q$ as 
        \begin{align}\nonumber
            r &= \sqrt{1-q(1-q)\epsilon^2}\\
            \theta &= \cos^{-1}(\sqrt{1-\epsilon^2(1-q)^2})
        \end{align}
  Hence, the matrix elements of the corresponding Jacobian read 
    	\begin{align} \nonumber
			\frac{\partial \epsilon}{\partial r}&=-\frac{r\epsilon}{1-r^2}\quad  &\frac{\partial \epsilon}
            {\partial \theta}&= \frac{\cos\theta}{1-q}\\
			\frac{\partial q}{\partial r}&=-\frac{2 r}{\epsilon^2(1-2q)} \quad   &\frac{\partial q}
            {\partial \theta}&=-\frac{\cos\theta}{\epsilon}\, ,
		\label{eq:derivativesrtheta}
	\end{align}
  and the resulting QFI matrix, up to leading order in $\epsilon$, is 
    	\begin{equation}
        \cF(\bm h(\boldsymbol \alpha))=\bematrix q(1-q) & -\epsilon q(1-2q)\\
                                            -\epsilon q(1-2q) & \frac{(1-2q)^2\epsilon^2}{1-q} \ematrix \,.
    	\label{eq:purity_angleQFI}
    	\end{equation}
   Observe that the matrix element associated with the precision in $q$ does not coincide with the optimal QFI~\cite{de_Vicente10, Rehacek17, Rehacek18} 
   indicating that trying to infer the relative intensity through purity and angle estimation does not yield optimal estimation precision.  That it does yield optimal precision 
   for the centroid---for all values of $q$---can be seen from the fact that, in our parametrization, $x_c=\sin\theta$ where $x_c$ is the position of the centre of 
   intensity for the two component mixture (see \figref{fig:2sources})\footnote{Recall that we have set the variance of the PSF to unity.  Specifically one has that $\frac{x_c}{\sigma}=\sin\theta$}.  
   The QFI matrix for estimating $\epsilon$ and $x_c$ reads, up to leading order in $\epsilon$,
        \begin{equation}
            \cF(\bm h(\boldsymbol \alpha))= \bematrix q(1-q) & 0\\ 0 & 1 \ematrix\, .
        \label{eq:QFI_centroid}
        \end{equation}
        
   Although, it is in principle possible to achieve maximum information on $x_c$, any measurement strategy derived from the initial state will probably depend on $x_c$, unless some sort of Haar 
   measurement process or feedback measurement is implemented.

\section{\label{sec:spectrum}The Schur-Sampling Measurement}
	
	Consider the state space $\cH_d^{\otimes N}$ of $N$ bosons with $d$ distinguishable degrees of freedom (modes).  This space carries the action of 
	two fundamental symmetry groups; the special unitary group of dimension $d$, $\mathrm{SU}(d)$, and the group of permutations of $N$ objects, $S_N$.
	The action of these two groups is mediated via the unitary representations $U^{\otimes N}:\mathrm{SU}(d)\to \mathbb{U}(\cH_d^{\otimes N})$ and 
	 $T: S_N\to\mathbb{U}(\mathcal{H}_d^{\otimes N})$ defined as  
		\be
			\begin{split}
				U^{\otimes N}_g	\, \ket{i_1,\ldots,i_N}&=\bigotimes_{n=1}^N U_g\,\ket{i_n}\,, \; \forall\, g\in\mathrm{SU}(d)\\
				T_\sigma\, \ket{i_1,\ldots,i_N}&=\ket{i_{\sigma(1)},\ldots,i_{\sigma(n)}}\,,\;\forall\, \sigma\in S_N\, ,
			\end{split}
		\label{eq:groupreps}
		\ee
	where $\{\ket{i_n}\, i_n\in (0,\ldots,d-1), \, \forall\, n\in(1,\ldots N)\}$ is an orthonormal basis of $\cH_d^{(n)}$. Both these representations are {\it compound} 
	meaning that they are composed of more fundamental \defn{irreducible} representations (or irreps for short).  Their decomposition into the latter can be obtained 
	via Schur's lemmas as 
		\be
    			\begin{split}
        			U_g^{\otimes N}&\cong\bigoplus_\mu U^{(\mu)}_g\otimes \one_\mu,\quad \forall\, g\in \mathrm{SU}(d)\\
        			T_\sigma&\cong\bigoplus_\nu T^{(\nu)}_\sigma\otimes \one_\nu, \quad \forall\, \sigma\in S_N\, 
    			\end{split}
    		\label{eq:reducibility}
		\ee
	where $\mu, \nu$ label the inequivalent irreps in the decompositions of $U^{\otimes N}$ and $T$ respectively, and are in one-to-one correspondence with 
	the \defn{conjugacy classes} of $\mathrm{SU}(d)$ and $S_N$ whilst $\one_{(\mu \slash \nu)}$ are identity matrices whose dimension equals to the number 
	of times the irreps $U^{(\mu)},\, T^{(\nu)}$ appear in their respective decompositions. The congruence in \eqnref{eq:reducibility} is understood as the 
	existence of an orthonormal basis relative to which the representations $U^{\otimes N}$ and $T$ assume their block-diagonal form.

\begin{figure}[ht!]  
  \centering
   \includegraphics[width=0.48\textwidth]{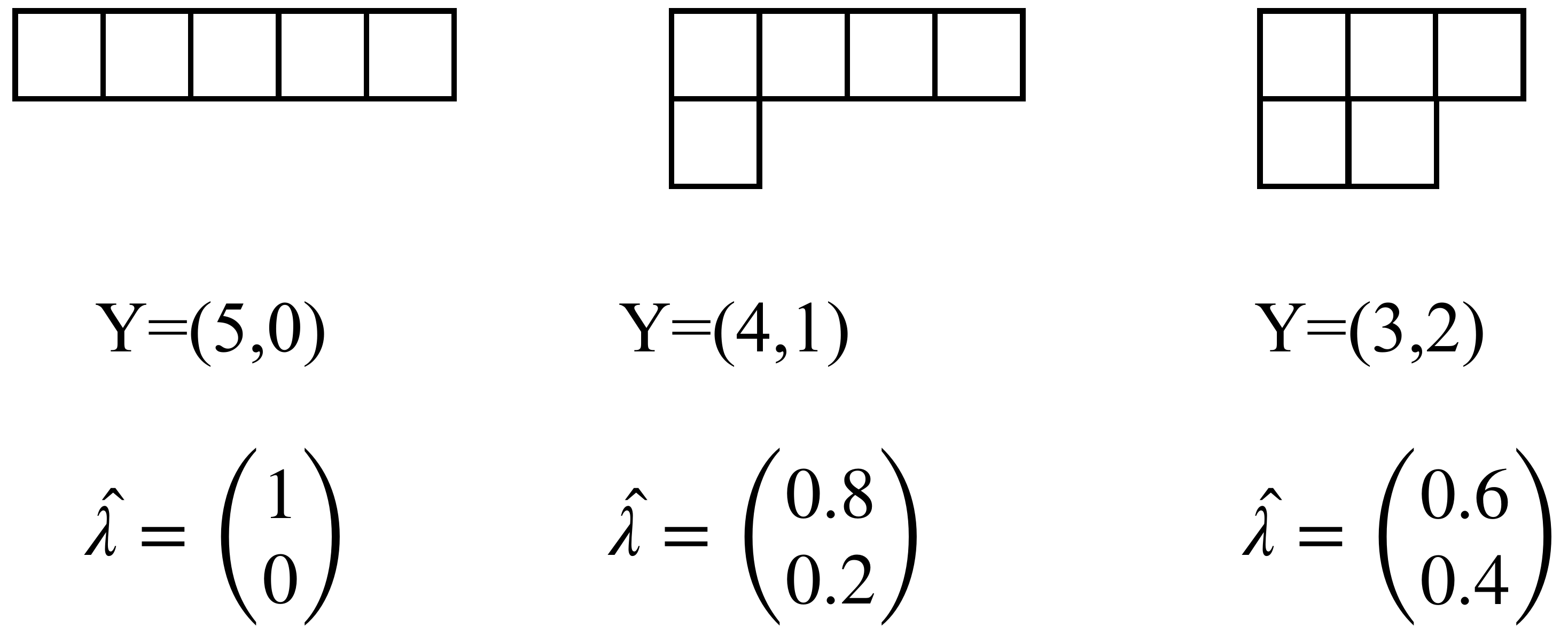}
    \caption{The Young frames of partitioning $5$ into at most two parts. The vectors $\hat{\bm \lambda}\in (0,1)^2;\, \hat{\bm\lambda}=\frac{Y}{N}$ are the corresponding estimates of the eigenvalues of $\rho\in\cB(\cH_2)$.}
   \label{fig:Young Frames}
\end{figure}

	As $[T_\sigma, U^{\otimes N}_g]=0, \; \forall\, \sigma\in S_N\, \mathrm{and}\, g\in \mathrm{SU}(d)$ it follows that 
	$[T^{(\nu)}_\sigma,U^{(\mu)}_g]=0$ for all $\mu,\nu$ and all $\sigma\in S_N, \, g\in\mathrm{SU}(d)$.  This implies that the representations $T$ 
	and $U^{\otimes N}$, as well as their irreps, are in each others {\it centralizer}. Making use of the double centralizer theorem Schur and Weyl proved that the 
	irrep labels $\mu, \nu$ are not independent but are in fact related~\cite{Weyl48}.  Indeed, the decompositions of both group actions can be labelled by either 
	the irrep labels of $\mathrm{SU}(d)$---related to the highest weight of the associated Lie algebra~\cite{Goodman09}---or those of $S_N$ which correspond to 
	the various ways one can partition the integer $N$ into at most $d$ parts in non-increasing order
		\be
   			\hspace{-0.5mm} Y:=\left\lbrace(Y_1,\ldots, Y_d)\,\mid\, Y_1\geq Y_2\geq\ldots\geq Y_d,\, \sum_{k=1}^dY_k=N\right\rbrace\,.
    		\label{eq;partitions}
		\ee
	The latter are known as \emph{Young frames} and are often visualized as an arrangement of $N$ boxes into at most $d$ rows such that the number of boxes in 	
	any row does not exceed those above it (see \figref{fig:Young Frames}).
	
	Schur-Weyl duality implies that there exists an orthonormal eigenbasis relative to which the total Hilbert space $\cH_d^{\otimes N}$ block diagonalizes as 
		\be
			\cH_d^{\otimes N}\cong\bigoplus_Y\, \cH_Y=\bigoplus_Y\, \cU_Y\otimes\cP_Y\,,
		\label{eq:blockdecomp}
		\ee
	where we have opted to use the irrep labels of $S_N$ to label the various invariant subspaces $\cH_Y$.  With respect to this basis the action of 
	representations $U^{\otimes N}$ and $T$ reads 
		\be
    			\begin{split}
        			U_g^{\otimes N}&\cong\bigoplus_Y U^{(Y)}_g\otimes \one,\quad \forall\, g\in \mathrm{SU}(d)\\
        			T_\sigma&\cong\bigoplus_Y \one\otimes T^{(Y)}_\sigma, \quad \forall\, \sigma\in S_N\, ,
			\end{split}
    		\label{eq:reducibility2}
		\ee
	and the subspaces $\cU_Y\subset\cH^{\otimes N}$ carry the irreps $U^{(Y)}$ of $\mathrm{SU}(d)$ whilst the subspaces $\cP_Y\subset\cH^{\otimes N}$ 
	carry the irreps $T^{(Y)}$ of $S_N$.  The orthonormal basis relative to which \eqnref{eq:blockdecomp} holds will be denoted as 
	$\{\ket{Y,M,x}\, : 1\leq M\leq\abs{\cU_Y}, \, 1\leq x\leq\abs{\cP_Y}\}$ and the transformation mapping the tensor product basis
	$\{\ket{i_1,\ldots,i_N}: 1\leq i_n\leq d,\, \forall \, n\}$ to the $\{\ket{Y,M,x}\}$ basis is the efficiently implementable Schur-Weyl transform~\cite{Bacon07}.
	For $d=2$ the integer partitions, $Y$, are in one-to-one correspondence with the total angular momentum of $N$ spin-1/2 systems
		\be
			J(Y)=\frac{Y_1-Y_2}{2},
		\label{eq:jvalue}
		\ee
	with $M$ the quantum number associated with the projection of angular momentum along some direction, $\hat{\bf n}$.
	
	The density operator $\rho^{\otimes N}\in\cB(\cH_2^{\otimes N})$, describing $N$ bosons each of which is given by the state $\rho$ of 
	\eqnref{eq:finite_mixture}, is invariant under any permutation of the $N$ bosons, i.e.,  
		\be
			\begin{split}
				\rho^{\otimes N}&=\frac{1}{N!}\sum_{\sigma\in S_N} T_\sigma \,\rho^{\otimes N}\,T^\dagger_\sigma\\
								&:=\cG_{S_N}[\rho^{\otimes N}],
			\end{split}
		\label{eq:SN-twirling}
		\ee
	where $\cG_{S_N}:\cB(\cH_2^{\otimes N})\to\cB(\cH_2^{\otimes N})$ is a completely positive, trace-preserving quantum operation 
	known as the permutation twirling map.  In terms of the decomposition of \eqnref{eq:blockdecomp} the latter map can be expressed as~\cite{Bartlett07}
		\be
			\cG_{S_N}\cong \sum_Y (\cI_Y\otimes \cD_Y)\circ \Pi_Y,
		\label{eq:G-twirling}
		\ee
	where $\Pi_Y$ are the projectors onto the subspaces $\cH_Y$ (see \eqnref{eq:blockdecomp}), 
	$\cD_Y:\cB(\cP_Y)\to\cB(\cP_Y),\, \cD[A]=\tr A\, \frac{\one}{\abs{\cP_Y}}$ is the completely depolarizing map, and 
	$\cI_Y:\cB(\cU_Y)\to\cB(\cU_Y)$ is the identity map.  Hence $\rho^{\otimes N}\in\cB(\cH_2^{\otimes N})$ takes the particularly simple form 
		\be
			\rho^{\otimes N}=\bigoplus_J p(J)\, \omega^{(J)}\otimes \frac{\one}{d_J},
		\label{eq:finite_mixtureblock}
		\ee
	where $\omega^{(J)}\in\cB(\cU_J)$ and $p(J)=\tr(\Pi_J \rho^{\otimes N})$ are given as~\cite{Bagan06}
		\be
			\begin{split}
				p(J)=&\left(\frac{1-r^2}{4}\right)^{\frac{N}{2}}d_J Z_J\\
				\omega^{(J)}=& \frac{1}{Z_J} \sum_{M=-J}^J R^M \ket{J,M}_{\hat{\bm r}}\bra{J,M}\\
				Z_J=&\sum_{M=-J}^J R^M=\frac{R^{J+1}-R^{-J}}{R-1}\\
				R=&\frac{1+r}{1-r}\, ,
			\end{split}
		\label{eq:pJtauJ}
		\ee
		\be
			d_J=\abs{\cP_J}={N\choose \frac{N}{2}-J}\frac{2J+1}{\frac{N}{2}+J+1}
		\label{eq:dimension_J}
		\ee
	and we have chosen to label the invariant blocks of the irrep decomposition according to their total angular momentum.  We note that the same decomposition
	holds for $N$ copies of the averaged state $\rho_{\mathrm{avg}}$ of \eqnref{eq:avg_state} with $r$ and ${\bf r}$ replaced with $r_{\mathrm{avg}}$ and 
	${\bf r}_{\mathrm{avg}}$ respectively.
	
	The form of \eqnref{eq:blockdecomp} makes it particularly intuitive to construct optimal estimation schemes pertaining to several intrinsic properties of $\rho$.  
	In estimating the spectrum, $\bm\lambda$, of a density operator $\rho\in\cB(\cH_d)$ the outcome $Y$ of the projective measurement $\{\Pi_J\}$ furnishes 
	the asymptotically efficient estimate, $\hat{\bm\lambda}=\frac{Y}{N}$~\cite{Keyl01}.  In the case of estimating the purity, $r$, of $\rho\in\cB(\cH_2)$, the 
	outcome $J(Y)$ of the projective measurement $\{\Pi_Y\}$ yields the asymptotically efficient estimate, $\hat{r}=\frac{2J(Y)}{N}$~\cite{Bagan05}.  Finally,   
	given multiple copies of two pure states $\ket{\psi},\ket{\phi}\in\cH_d$ the measurement outcomes of $\{\Pi_Y\}$ are known to provide an efficient estimator,
	$\hat{c}=\left(\frac{2J(Y)}{M+N}\right)^2$, of the overlap between two states $\ket{\psi}, \,\ket{\phi}$, where $M$, $N$ 
	are the number of copies of the two states respectively~\cite{Fanizza20}.  
	
	For $N=2$, there are only two Young frames associated with the symmetric and anti-symmetric subspaces of $\cH_d^{\otimes 2}$. The corresponding projection
	operators are given by 
		\be
			\begin{split}
				\Pi_{Y=S}&=\frac{\one +\mathrm{SWAP}}{2}\\
				\Pi_{Y=A}&=\frac{\one -\mathrm{SWAP}}{2}\,
			\end{split}
		\label{eq:SWAPtest}
		\ee	
	where $\mathrm{SWAP}:\cH_d^{\otimes 2}\to\cH_d^{\otimes 2}$ is the operator defined as 
		\be
			\mathrm{SWAP}\ket{i}\otimes\ket{j}=\ket{j}\otimes\ket{i}, \forall, \ket{i},\ket{j}\in\cH_d\, .
		\label{eq:SWAP}
		\ee
	Given two pure states $\ket{\psi},\ket{\phi}\in\cH_d$, the probability of obtaining their symmetric or anti-symmetric components is given by 
		\begin{equation}
			\begin{split}
				P_S&=\frac{1+\lvert\inner{\psi}{\phi}\rvert^2}{2}\\
				P_A&=\frac{1-\lvert\inner{\psi}{\phi}\rvert^2}{2}\, .
			\end{split}
		\label{eq:SAprobs}
	\end{equation}
	Given $N$ copies of each of the two states, performing the measurement of \eqnref{eq:SWAPtest} on each pair, and registering the number of times  
	the symmetric (equiv. anti-symmetric) outcome is obtained one can extract an estimate of the overlap $\abs{\inner{\psi}{\phi}}$.  Indeed, this 
	protocol---known as the SWAP test---is of fundamental importance in quantum cryptography~\cite{Buhrman01} and quantum information~\cite{Ekert02}.  
	
	If the states in question describe the degrees of freedom of bosons then there exists a simple experimental implementation of the measurement of 
	\eqnref{eq:SWAPtest} by means of the Hong-Ou-Mandel interference effect~\cite{GarciaEscartin13}. If both bosons encode the same signal, then they will bunch 
	resulting in only a single detector clicking with certainty.  However, if the bosons encode different signals, then there is a non-zero probability that both detectors
	click, resulting in a coincidence count.  By repeating the experiment a sufficiently large number of times and registering the number of coincidence counts one can 
	extract information about the separation, $\epsilon$, between the two signals~\footnote{In~\cite{Parniak18, Ansari21, Mazelanik22}, the signals were assumed to
	have equal weight and the parameters of interest where the spatial, spectral, or temporal separation of the signals as well as their centre of intensity}. 
	
	However, it is known that the SWAP test is not the most efficient way to extract precise estimates of the overlap~\cite{Badescu19, Fanizza20}.  Indeed, whilst 
	super-resolution is possible by processing bosons in pairs, one only achieves a fraction of the ultimate quantum limit.  
	
\end{document}